

\magnification\magstep1
\input degmacs
\pageno=1

\runningheads{A path-integral approach to surgical
invariants of three-manifolds}{Bogus\l aw Broda}

\def\ml{{{\cal M}_L}}
\def\sthree{{{\cal S}^3}}
\def\tr{{\rm Tr}}
\def\half{{1\over2}}
\def\hol#1{{\rm Hol}_{K_{#1}}(A)}
\def\cd{{\cal D}}
\def\co{{\cal O}}
\def\wilson#1{W_{\mu_{#1}}^{K_{#1}}(A)}
\def\hg{\hat G}
\def\prodin{\prod_{i=1}^N}

\def\sigmami#1{\sum_{\mu_{#1}\in\hg}}
\def\partition{Z(\sthree,L;g_1,g_2,\dots,g_N)}
\def\avewilson#1#2{\left<W_{#1}^{#2}(A)\right>}
\def\sigmaad{\sum_{a=1}^d}
\def\tensorpr#1#2{t_{#1}^a\otimes t_{#2}^a}

\def\one{{\bf 1}}


\def\plusskein{
\setbox0=\hbox{
\setbox1=\hbox{\bigg/}
\setbox2=\hbox{$\backslash$}
\hbox to -\wd1{} \copy1
\kern-\wd1\raise.5\ht1\copy2
\lower\dp1\hbox{\raise\dp2\copy2}
}\copy0}

\def\minusskein{
\setbox0=\hbox{
\setbox1=\hbox{$\bigg\backslash$}
\setbox2=\hbox{/}
\hbox to -\wd1{} \copy1
\kern-\wd1\lower.5\ht1\copy2
\raise\dp1\hbox{\lower\dp2\copy2}
}\copy0}

\def\zeroskein{
\;\bigg\vert\;\bigg\vert\;}

\def\nullskein#1{
\;\bigg\vert #1\;}

\vbox to 2cm
{\noindent
$^{\rm
Subm.~to~Proc.:~2nd~Gauss~Symp.,~1993,
{}~eds.~R.~Fritsch,~Gruyter~Publ.~Co.
\quad\bf
hep-th/9312116
}$}
\vskip-2cm

\Chapter
Chern-Simons approach to three-manifold invariants

\Author
Bogus\l aw Broda*

\Abstract
A new, formal, non-combinatorial approach to invariants of
three-dimensional manifolds of Reshetikhin, Turaev and
Witten in the framework of non-perturbative topological
quantum Chern-Simons theory, corresponding to an arbitrary
compact simple Lie group, is presented. A direct
implementation of surgery instructions in the context of
quantum field theory is proposed. An explicit form of the
specialization of the invariant to the group SU(2) is
derived.

\skipaline
\Section 1. Introduction

\vfootnote{*}
{The work has been supported by the Alexander von Humboldt
Foundation and the KBN grant 202189101.}%
In 1989 Edward Witten proposed a new interesting
topological invariant of {\it three-dimensional\/}
manifolds in his famous paper on quantum field theory and
the Jones polynomial [1]. An explicit construction of the
invariant, using quantum groups, appeared for the first
time in a paper of Reshetikhin and Turaev [2]. Other papers
presenting re-derivations of this result are more
geometrical by nature [3], and use the Temperley-Lieb
algebra, as suggested by Lickorish [4--6]. All the
approaches are combinatorial. Non-combinatorial
possibilities, rather straightforward though mathematically
less rigorous, are offered by {\it topological} quantum
field theory (see, e.~g. [7]).

Inspired by the papers [6] and [8], we aim to propose a
new, formal, non-combinatorial derivation of the
three-manifold invariants of Reshetikhin, Turaev and Witten
(RTW) in the framework of non-perturbative (topological)
quantum Chern-Simons (CS) gauge theory. The idea is
extremely simple, and in principle applies to an arbitrary
compact (semi-)simple group $G$ (not only to the SU(2)
one). Our invariant is essentially the partition function
of CS theory on the manifold $\ml$ defined via {\it
surgery} on the framed link $L$ in the three-dimensional
sphere $\sthree$.  Actually, surgery instructions are
implemented in the most direct and literal way. The method
of cutting and pasting back, which has been successfully
applied to two-dimensional Yang-Mills theory [8], is
explicitly used in the standard field-theoretical fashion.
Roughly speaking, cutting corresponds to fixing, whereas
pasting back to identification and summing up the boundary
conditions.

As a by-product of our analysis we consider the {\it
satellite formula}, and derive the {\it Kauffman bracket}
polynomial invariant of a trivial (with zero framing)
unknot for an arbitrary representation of SU(2).
\Section 2. General formalism

Our principal goal is to compute the partition function
$Z(\ml)$ of CS theory on the manifold $\ml$ defined via
(honest/integer) surgery on the framed link
$L=\bigcup_{i=1}^NK_i$ in $\sthree$, for an arbitrary
compact simple (gauge) Lie group $G$. Obviously, the
starting point is the partition function of CS theory
$Z(\sthree)$ on the sphere $\sthree$ [1]
$$
Z(\sthree)=\int e^{ik{\rm cs}(A)}{\cal D}A,
\eqno(2.1)
$$
where the functional integration is performed with respect
to the connections $A$ modulo gauge transformations,
defined on a trivial $G$ bundle on $\sthree$, and
$k$ is the level ($k\in\bf Z^+$). The classical action is
the {\it CS secondary characteristic class}
$$
{\rm cs}(A)={1\over4\pi}\int_\sthree\tr\left(AdA
+{2\over3}A^3\right),
\eqno(2.2)
$$
and the expectation value of
an observable $\co$ is defined as
$$
\left<\co\right>=\int\co
e^{ik{\rm cs}(A)}\cd A.
\eqno(2.3)
$$

According to the surgery prescription we should cut out a
closed tubular neighbourhood $N_i$ of $K_i$ (a solid
torus), and paste back a copy of a solid torus $T$,
matching the meridian of $T$ to the (twisted by framing
number) longitude on the boundary torus $\partial N_i$ in
$\sthree$ [5,~9]. To this end, in the first step, we should
fix boundary conditions for the field $A$ on the twisted
longitude represented by $K_i$.  Since the only
gauge-invariant (modulo conjugation) quantity defined on a
closed curve is holonomy [8], we associate the {\it
holonomy} operator $\hol{i}$ to each knot $K_i$.  Thus the
symbol
$$
\partition\eqno(2.4)
$$
should be understood as the {\it constrained} partition
function of CS theory, i.~e. the values of holonomies along
$K_i$ are fixed
$$
\hol{i}=g_i,\qquad i=1,2,\dots,N.
\eqno(2.5)
$$
Now, we can put
$$
\partition
=\left<\prodin\delta(g_i,\hol{i})\right>,
\eqno(2.6)
$$
where $\delta$ is a (group-theoretic) Dirac delta-function
[8]. Its explicit form following from the (group-theoretic)
Fourier expansion [10] is
$$
\delta(g,h)=\sigmami{}\overline{\chi_\mu(g)}\chi_\mu(h).
\eqno(2.7)
$$
Physical observables being used in CS theory are typically
Wilson loops, defined as
$$
\wilson{}=\tr_\mu(\hol{})\equiv\chi_\mu(\hol{}),
\eqno(2.8)
$$
where $\mu\in\hg$ numbers inequivalent irreducible
representations (irrep's) of $G$, and $\chi_\mu$ is a
character. Effectively, the set $\hg$ is reduced under the
symbol ``$\left<~\right>$''. By virtue of (2.7--8)
$$
\delta(g_i,\hol{i})
=\sigmami{}\overline{\chi_\mu(g_i)}W_\mu^{K_i}(A).
\eqno(2.9)
$$
Inserting (2.9) into (2.6) yields, as a basic building
block, the following representation of the constrained
partition function
$$
\partition=\left<\prodin\sigmami{i}
\overline{\chi_{\mu_i}(g_i)}\wilson{i}\right>.
\eqno(2.10)
$$
In the second step of our construction, we should paste
back the tori matching the pairs of ``longitudes'' (the
twisted longitudes and the meridians), i.~e. we should
identify and sum up the boundary conditions. Since the
interior of a solid torus is homeomorphic to $\sthree$ with
a removed solid torus, actually the meridians play the role
of longitudes in analogous cutting procedures for an unknot
$\left\{\bigcirc\right\}$ (with reversed orientation).
Thus the partition function of CS theory on $\ml$ is
$$
Z(\ml)={1\over N_L}\int\prodin dg_i Z(\sthree,\bigcirc;g_i^{-1})
\partition\qquad\qquad\qquad\quad
$$
$$
\qquad\qquad
={1\over N_L}\int\prodin dg_i\sigmami{i}\sum_{\nu_i\in\hg}
\overline{\chi_{\mu_i}(g_i^{-1})} \,
\overline{\chi_{\nu_i}(g_i)}
\avewilson{\mu_i}{\bigcirc}
\left<\prod_{j=1}^N W_{\nu_j}^{K_j}(A)\right>,
\eqno(2.11)
$$
where $N_L$ is a link-dependent normalization constant, and
the reversed orientation of the unknots
$\left\{\bigcirc\right\}$ (corresponding to the meridians
of the pasted back tori) accounts for the power $-1$ of the
group elements $g_i$. From the orthogonality relations for
characters and unitarity of irrep's, it follows that the
three-manifold invariant is of the form
$$
Z(\ml)={1\over
N_L}\left<\prodin\omega_{K_i}(A)\right>,
\eqno(2.12{\rm a})
$$
where
$$
\omega_{K_i}(A)\equiv\sigmami{i}\avewilson{\mu_i}{\bigcirc}
\wilson{i}
\eqno(2.13)
$$
is an element of the linear skein of an annulus, immersed
in the plane as a regular neighbourhood of $K_i$ [6].
$\langle W_\mu^\bigcirc(A)\rangle$ are some computable
coefficients depending on $\mu$, $k$ and $G$. Eq.~(2.12a)
can be easily generalized to accommodate an ordinary link
${\cal L}=\bigcup_{i=1}^M{\cal K}_i$ embedded in $\ml$
$$
\left<\prod_{i=1}^M W_{\mu_i}^{{\cal K}_i}(A)\right>_\ml
={1\over N_L}\left<\prod_{i=1}^M W_{\mu_i}^{{\cal K}_i}(A)
\prod_{j=1}^N\omega_{K_j}(A)\right>.
\eqno(2.12{\rm b})
$$
We defer the solution of the issue of the determination of
the normalization constant $N_L$ to the end of Sect.~4.
\Section 3. The satellite formula

The easiest way to calculate $\avewilson{\mu}{\bigcirc}$
follows from the satellite formula [11,~12]. In turn, the
simplest derivation of the satellite formula on the level
of skein relations, in the context of topological field
theory, could look as follows. Let us consider the
topological-field-theory approach to skein relations, which
yields the \hbox{(quasi-)}braiding matrix $B$ in the form
[12,~13]
$$
B=q^{\sigmaad\tensorpr{\mu}{\nu}},
\eqno(3.1)
$$
where
$$
q=e^{-{2\pi i\over k}},
\eqno(3.2)
$$
and $\mu$, $\nu$ are two irrep's of the $d$-dimensional
group $G$.  The square of $B$, the monodromy matrix $M$
($M=B^2$) can be derived, for example, in the framework of
the path-integral approach to link invariants (advocated in
[13]) as the contribution coming from the intersection of
the surface $\cal S$ corresponding to the representation
$\mu$ and the line $\ell$ corresponding to $\nu$. If we
double the line $\ell$, possibly assigning different
representations to each of the components, say $\nu$ and
$\lambda$, there will appear two intersection points and
consequently two contributions giving rise to
$$
B=q^{\sigmaad\tensorpr{\mu}{\nu}}\,
q^{\sigmaad\tensorpr{\mu}{\lambda}}\qquad
$$
$$
\qquad=q^{\sigmaad(\tensorpr{\mu}{\nu}\otimes\one_{\lambda}
+t_\mu^a\otimes\one_\nu\otimes t_\lambda^a)}
=q^{\sigmaad\tensorpr{\mu}{\nu\otimes\lambda}},
\eqno(3.3)
$$
where
$$
t_{\nu\otimes\lambda}^a\equiv
t_\nu^a\otimes\one_\lambda+\one_\nu\otimes
t_\lambda^a
\eqno(3.4)
$$
is a generator of $G$ in the product representation
$\nu\otimes\lambda$. Hence we have the satellite formula
(at least on the level of skein relations)
$$
W_\mu^{\cal K}(A) W_\nu^{\cal K}(A)
\approx W_{\mu\otimes\nu}^{\cal K}(A),
\eqno(3.5)
$$
where ``$\approx$'' means the ``weak equality'',
$$
X\approx Y \Longleftrightarrow
\langle X\rangle=\langle Y\rangle.
\eqno(3.6)
$$
The product on LHS of (3.5) should be understood in a
``regularized'' form, i.~e. the both $\cal K$'s should be
split up. Obviously, Eq.~(3.5) can be readily
generalized by induction to any number of factors, whereas
RHS of (3.5) can be expanded into irreducible components
of the product $\mu\otimes\nu$.
\Section 4. SU(2)-invariant

In this section, we derive an explicit form of the
specialization of our invariant (2.12) to the group SU(2),
and show that it agrees with the result of Lickorish [6].

It appears that a very convenient way of organization of
irrep's of SU(2) group is provided by the polynomials
$S_n(x)$, closely related to the Chebyshev polynomials.
$S_n(x)$ are defined recursively by the formula
$$
S_{n+2}(x)=xS_{n+1}-S_n(x),\qquad n=0,1,\dots,
\eqno(4.1{\rm a})
$$
together with the initial conditions
$$
S_0(x)=1,\qquad S_1(x)=x.
\eqno(4.1{\rm b})
$$
Usefulness of $S_n(x)$, in the context of the SU(2) RTW
invariant, has been observed in [4,~6]. By virtue of the
definition (4.1), $S_n(x)$ expresses $n$-th irrep of
SU(2) in terms of powers of the fundamental representation
$x$, denoted as $\one$ henceforth. The explicit solution of
(4.1) is
$$
S_n(2\cos\alpha)={\sin((n+1)\alpha)\over\sin\alpha}.
\eqno(4.2)
$$

For the group SU(2) the satellite formula (3.5) now assumes
the following elegant form
$$
W_n^{\cal K}(A)=W_{S_n(\one)}^{\cal K}(A)
\approx S_n\left(W_\one^{\cal K}(A)\right),
\eqno(4.3)
$$
whereas the skein relations for the fundamental
representation ($n=1$) of SU(2)
$$
q^{1\over4}\left<\left\{\plusskein\right\}\right>
-q^{-{1\over4}}\left<\left\{\minusskein\right\}\right>
=(q^{\half}-q^{-\half})
\left<\left\{\zeroskein\right\}\right>,
\eqno(4.4{\rm a})
$$
$$
\left<\left\{\nullskein{\pm1}\right\}\right>
=-q^{\pm{3\over4}}\left<\left\{\nullskein{0}\right\}\right>,
\eqno(4.4{\rm b})
$$
where the integers in (4.4b) mean a framing. Closing
the left legs of all the (three) diagrams in (4.4a) with
arcs, as well as the right ones, next applying (4.4b), and
using the property of locality, we obtain
$$
-(q-q^{-1})\avewilson{\one}{\bigcirc}
=(q^{\half}-q^{-\half})\avewilson{\one}{\bigcirc\bigcirc}
=(q^{\half}-q^{-\half})\avewilson{\one}{\bigcirc}^2.
\eqno(4.5)
$$
Hence
$$
\avewilson{\one}{\bigcirc}=-\left(q^{\half}+q^{-\half}\right)
=-2\cos{\pi\over k},
\eqno(4.6)
$$
and by virtue of the satellite formula (4.3)
$$
\avewilson{n}{\bigcirc}=S_n\left(-2\cos{\pi\over k}\right)
=(-)^n{\sin{(n+1)\pi\over k}\over\sin{\pi\over k}}
=(-)^n{q^{n+1\over2}-q^{-{n+1\over2}}\over
q^{\half}-q^{-\half}}.
\eqno(4.7)
$$
We can observe a remarkable property of (4.7) for
$n=k-1$, limiting the set $\hg$, namely
$$
\avewilson{k-1}{\bigcirc}=0.
\eqno(4.8)
$$
It appears that for any $\cal K$
$$
\langle\cdots W_{k-1}^{\cal K}(A)\cdots\rangle=0.
\eqno(4.9)
$$
Actually, we are dealing with a tensor algebra of finite
order, which can be interpreted as a fusion algebra [14].
In particular, Eq.~(4.9) immediately follows from (4.8)
for any $\cal K$ that can be unknotted with corresponding
skein relations.  Thus we can truncate representations
of SU(2) above the value $k-2$, and assume
$$
0\leq n\leq k-2,\qquad k=2,3,\dots.
\eqno(4.10)
$$
The final explicit form of $\omega_K$ for
the group SU(2) is then
$$
\omega_K(A)=\sum_{n=0}^{k-2}(-)^n
{q^{n+1\over2}-q^{-{n+1\over2}}\over
q^{\half}-q^{-\half}}S_n\left(W_\one^K(A)\right),
\eqno(4.11)
$$
and agrees with the corresponding expression derived by
Lickorish with a combinatorial method [6]. Strictly
speaking, $Z(\ml)$ is invariant with respect to the second
Kirby move $K_2$. It means that it is insensitive to the
operation of sliding one of its handles over another one.
But up to now we have not considered the issue of the
determination of the normalization constant $N_L$. It
appears that proper normalization of the partition function
$Z(\ml)$ universally follows from the requirement of its
invariance with respect to the first Kirby move $K_1$.
Hence the normalization constant $N_L$ can be chosen in the
form [6]
$$
N_L=
\left<\omega_{\bigcirc_{+1}}(A)\right>^{b_+(L)}
\left<\omega_{\bigcirc_{-1}}(A)\right>^{b_-(L)},
\eqno(4.12)
$$
where $b_+(L)$ ($b_-(L)$) is the number of positive
(negative) eigenvalues of the linking matrix of $L$.

\References 10

\ref 1  Witten, E., Commun. Math. Phys. 121 (1989), 351
\ref 2  Reshetikhin, N. and Turaev, V. G., Invent.
math. 103 (1991), 547
\ref 3  Kirby, R. and Melvin, P., Invent. math. 105 (1991),
473
\ref 4  Lickorish, W. B. R., Math. Ann. 290 (1991), 657
\ref 5  Kauffman, L. H., Knots and Physics. World
Scientific, Singapore 1991
\ref 6  Lickorish, W. B. R., The skein method for
three-manifold invariants. Preprint and talk given in the
framework of the programme on Low Dimensional Topology
and Quantum Field Theory, Cambridge 1992
\ref 7  Guadagnini, E., Nucl. Phys. B 375 (1992), 381
\ref 8  Blau, M. and Thompson, G., Int. J. Mod.
Phys. A 7 (1992), 3781
\ref 9  Rolfsen, D., Knots and Links. Publish or
Perish, Wilmington 1976
\ref 10  Wallach, N. R., Harmonic analysis on homogeneous
spaces. Marcel Dekker, 1973
\ref 11  Morton, H. R. and Strickland, P., Math. Proc.
Camb. Phil. Soc. 109 (1991), 83
\ref 12  Guadagnini, E., Int. J. Mod. Phys. A 7 (1992), 877
\ref 13  Broda, B., Mod. Phys. Lett. A 5 (1990), 2747.
Phys. Lett. B 271 (1991), 116
\ref 14  Guadagnini, E., Phys. Lett. B 260 (1991), 353

\vfill

\noindent
Bogus\l aw Broda, Department of Theoretical Physics,
University of \L\'od\'z, Pomorska 149/153, PL--90-236
\L\'od\'z, Poland

\noindent
e-mail: bobroda@plunlo51.bitnet

\bye